# IONIZATION BY IMPACT ELECTRONS IN SOLIDS: ELECTRON MEAN FREE PATH FITTED OVER A WIDE ENERGY RANGE

## Beata Ziaja

*HASYLAB at DESY, Notkestr. 85, D-22607 HAMBURG, Germany*

*and*

*Department of Theoretical Physics, Institute of Nuclear Physics, Radzikowskiego 152, 31-342 Cracow, Poland*

## Richard A. London [1]

*Lawrence Livermore National Laboratory, Livermore, CA 94551, USA*

## Janos Hajdu

*ICM Molecular Biophysics, Biomedical Centre, Uppsala University, Husargatan 3, Box 596, S-75123 Uppsala, Sweden*

---

**Abstract:**
We propose a simple formula for fitting the electron mean free paths in solids both at high and at low electron energies. The free-electron-gas approximation used for predicting electron mean free paths is no longer valid at low energies ($E < 50$ eV), as the band structure effects become significant at those energies. Therefore we include the results of the band structure calculations in our fit. Finally, we apply the fit to 9 elements and 2 compounds.

# 1   Introduction

Inelastic interactions of electrons with solids [1] are of a great importance for several measurement techniques [2] including energy loss spectroscopy, low energy electron diffraction, photoemission spectroscopy and time-resolved two-photon photoemission. At present electron mean free paths are well known in two energy regimes: at high energies, where they are predicted either with Bethe(-like) equations [3,4] or with the more accurate optical models based on the free-electron-gas approximation [5,6], and at very low energies, where they are calculated either with experimental data on electron lifetimes or with first-principles calculations [2,7–20]. In the low energy region electron mean free paths have been extensively studied, especially in semiconductors, where they are needed to understand the properties of semiconductor devices under high electric fields (see e.g. [14]).

Accurate calculation of the electron mean free path for impact ionizations is essential for investigating the radiation damage by energetic photons in solids. With the anticipated developement of free-electron lasers, damage to solid materials caused by an intense X-ray irradiation has become of significant interest to the research community. Radiation damage is the limiting factor in the achievable resolution for biological materials in X-ray diffraction as well as in electron microscopy [21–23]. New X-ray sources, like free electron lasers (XFELs), will soon provide very short, intense pulses that may allow existing damage limitations to be overcome [24]. A fundamental understanding of the interaction of X-rays with solid state materials is important to pursue this possibility. Damage is also a limiting factor in the design of X-ray optics [25] and detectors for XFELs and for the survival of samples exposed to the intense X-ray beam. On the positive side, production of "warm dense matter" [26] by XFELs will be mediated by electron cascades similar to those that underlie the damage processes. Also, interpretation of recent experiments on non-thermal melting in solids [27] and some unexpected behavior of xenon clusters exposed to very short UV pulses [28] depend on an understanding of electron cascades. Soon it will be possible to study the time dependent electron cascades experimentally, with the XFELs.

X-rays interact with the material mainly via the photoelectric effect. In light elements, the emission of an energetic photoelectron is predominantly followed by the emission of a less energetic Auger electron [24]. These electrons propagate through the sample, and cause further damage by excitations of secondary electrons. The extent of ionisation will depend on the size of the sample. Photoelectrons released by X-rays of $\sim 1$ Å wavelength are fast, $v \sim 660$ Å/fs, and they can escape from small samples early in an exposure. In contrast, Auger electrons are slow ($v \sim 95$ Å/fs in carbon), so they remain longer in a sample, and it is likely that they will thermalize there. A detailed description of electron cascades initiated by an electron impact is needed for a better understanding of radiation damage especially in larger samples as secondary ionization caused by propagating photoelectrons becomes significant there.

The electron transport models constructed to date for studying interactions of short pulse X-rays with matter (see e.g. Refs. [29,30]) have been based on mean free paths for impact ionization obtained with optical models [5,6]. Those mean free paths were valid at high energies only, $E > 50$ eV. Their extension to lower energies was questionable, and it lead to the underestimation of the total number of secondary electrons released, as the electrons liberated by carriers of low impact energies were then neglected in the simulations. A need for a unified model that can extend electron mean free paths down to very low impact energies was expressed in Ref. [30]. An extension of the TPP-2 optical model [5] down to the very low energies, $E = 5.47$ eV, was proposed there. Simulations applying the model extended were performed in diamond, and they yielded a reasonable estimate of the late number of secondary electrons, released by a single impact electron.

Here we advance previous work by developing a simple formula for fitting the electron mean free paths for impact ionization in various solids over a wide energy range. The accuracy of the fit obtained with this model is sufficient for a correct calculation of the number of secondary ionizations induced by an impact electron. In this paper, we first give the theoretical foundations of the model. Then, the model is applied to fit electron mean free paths in 9 elements and 2 compounds. Both metals and semiconductors are considered. The accuracy of the fit at different energy regimes is then discussed. Possible applications and extensions of the fit are proposed. Finally, a simple approximation for the differential cross sections is suggested, which, together with our fit to the mean free paths may be used to construct a simple and computationally efficient model of the electron transport.

Although the electron transport in solids has been extensively studied using different techniques, up to our knowledge, no universal fit of the electron mean free paths working both at high and at low energies and including results of the first-principles calculations, has been yet published. We hope that our result will help to fill up this void.

## 2 Theory

### 2.1 Physical picture

Electrons propagating in a solid interact with the atoms of the solid. These interactions may be either inelastic or elastic. The average distance travelled by an electron between two consecutive inelastic or elastic collisions is described by a mean free path, $\lambda_{in(el)}$, which is proportional to the inverse of the inelastic (elastic) cross section.

During an inelastic collision the impact electron loses a part of its energy, transferring it to another electron(s) or to the lattice. There are several energy loss channels accessible for the primary electron: (i) direct production of electron-hole pairs in core or valence ionizations, (ii) collective excitations of the solid (plasmons, excitons and phonons), [31–33]. The accessibility of the loss channels depends on the impact (kinetic) energy of the primary electron, $E$. If $E$ is larger than the threshold for the core ionizations, $E_{CORE}$, all routes of the energy loss are available for the primary electron, and the direct production of electron-hole pairs is the most likely route. If the energy of the primary electron is lower than $E_{CORE}$ but still higher than the band gap width, $E_G$, which is the minimal energy needed to create an electron-hole pair, ionizations of the valence band can proceed. Pair production remains a dominating mechanism of the energy loss.

If the energy of the primary impact lies below the threshold for plasmon excitations, $E < E_P$ ($E_P < E_{CORE}$), the plasmonic channel of energy loss also closes. At even lower energies, the production of electron-hole pairs will be suppressed. This occurs at energies, $E < E_G$. For metals we have, $E_G \simeq 0$, so that the direct production of electron-hole pairs in metals will be possible at any non-zero impact energy, $E > 0$.

In semiconductors and insulators long-living excitations of bound electron-hole pairs called excitons are also possible. The thresholds for excitons lie below $E_G$. At energies below exciton thresholds only phonon excitations occur with the energy gains or losses of less than $0.1$ eV. Phonons may be excited also in metals. Due to the small energy transfer required for a phonon exchange, the phononic channel of energy loss remains open even at very low impact energies both in metals and in the semiconductors.

As we are interested in calculating the mean free path of an impact electron between two consecutive impact ionization events, $\lambda_{ii}$, we restrict our analysis to processes which contribute significantly to the excitations of secondary electrons. These are the processes of : (i) the direct pair production allowed at energies, $E > E_G$, and (ii) the plasmon excitations allowed at energies, $E > E_P$, where $E_P > E_G$. In case of plasmonic excitations the secondary electrons are produced indirectly, as a result of the plasmon decay [31]. Although, with the free-electron-gas model of solids [11,31], only volume plasmons of large wave vectors can decay into single electron-hole pairs, in real solids volume plasmons of even very small wave vectors may decay into electron-hole pairs via interband transitions [33]. Therefore one may roughly assume that every plasmon excitation produces a pair of a secondary electron and a hole.

## 2.2 Impact ionization rate, $\Gamma(E)$, at low energies, $E_G < E < E_P$

Electrons of low energies, $E_G < E < E_P$, moving in solids experience strong electron-electron scattering processes. In metals, inelastic lifetimes of these electrons, $\tau$, have been studied for many years in the framework of the free-electron-gas model of solids [31, 32]. However, recent experiments and calculations [2, 11] have shown that the band structure of metals is very important for the electron transport at low electron energies. The calculations have shown that the lifetimes of the electrons are strongly affected by the topology of the Fermi surface and the density of states in bands, even for free-electron-like metals for which the free-electron-gas model works accurately (e.g. aluminium, see [7, 9, 32]).

Strong dependence of electron scattering on the details of the band structure was also observed in semiconductors and insulators. In these solids, the transport of hot electrons has been intensively studied, both theoretically and experimentally, as its correct modelling is essential in order to describe the transport of carriers in semiconductor devices [14].

The interaction of an impact electron with the electronic band of the solid, which may lead to an impact ionization, proceeds via the screened Coulomb potential [2, 7, 15]. Different first-principle formalisms may be applied in order to describe this interaction. In metals, the scattering rate of an impact electron, $\Gamma = 1/\tau$, is usually obtained with many-body theory from the imaginary part of the electron self-energy [2, 7, 8, 10]. The rate, $\Gamma_{n,k}$, is a product of the Fourier transform of the screened Coulomb interaction and of the expansion coefficients of the electron wave functions sampled in the irreducible part of the Brillouin zone [2, 7]. This rate depends on the wave vector, $k$ and the energy band, $n$. Therefore it has to be summed over all $k$'s and $n$'s available at the same (impact) energy in the Brillouin zone in order to obtain the average value, $\Gamma(E)$, as a function of the impact energy.

In semiconductors, the impact ionization rate is usually calculated with the scattering theory, using time-dependent perturbation expansion [9, 13, 15–20]. Applying the Fermi golden sum rule, the $1^{st}$ order contribution to the ionization rate sums the elements of the scattering matrix, $M^2(1, 2, 3, 4)$, which describe the screened Coulomb interaction of the primary electron (index 1) with the valence band, resulting in a release of a secondary electron (index 3) and a secondary hole (index 4) [13]. Index 2 denotes the final state of the primary electron. These matrix elements are multiplied by Dirac $\delta$-functions which impose the energy and momentum conservation. The set of matrix elements is then summed over all energy bands and wave vectors available in the first Brillouin zone,

$$\Gamma(E) = c \sum_{n_1,n_2,n_3,n_4} \int d^3k_1 \ldots d^3k_4\ \delta(\epsilon_{n_1}(\mathbf{k_1}) - E)\ \delta(\epsilon_{n_1}(\mathbf{k_1}) + \epsilon_{n_4}(\mathbf{k_4}) - \epsilon_{n_2}(\mathbf{k_2}) - \epsilon_{n_3}(\mathbf{k_3}))$$
$$\times\ \delta(\mathbf{k_1} + \mathbf{k_4} - \mathbf{k_2} - \mathbf{k_3} - \mathbf{K_0})\ M^2(1,2,3,4)/\sum_{n_1}\delta(\epsilon_{n_1}(\mathbf{k_1}) - E), \quad (1)$$

where c is a normalization constant, $M^2(1,2,3,4) = M^2(\mathbf{k_1}, n_1; \mathbf{k_2}, n_2; \mathbf{k_3}, n_3; \mathbf{k_4}, n_4)$ sums the direct and the exchange interaction terms and their interference [13]. The shape of the energy function, $\epsilon_n(\mathbf{k})$, follows from the dispersion relation in the nth band. The vector $K_0$ is a principal lattice vector, introduced in order to ensure that $\mathbf{k_1}, \mathbf{k_2}, \mathbf{k_3}, \mathbf{k_4}$ are all in the first Brillouin zone.

The self-energy approach and the approach based on scattering theory may be applied both for metals and for semiconductors. We classified them as the "metal" approach and the "semiconductor" approach as the majority of the papers treating the electron transport in metals uses the self-energy formalism for their band calculations, and the scattering theory approach is commonly used by the semiconductor community. Alternative treatments, using the self-energy approach for semiconductors and the scattering theory for metals can also be found, see e. g. Refs. [9, 12, 13].

Finally, we note that first-principles calculations are always complex and difficult. They require detailed information on the structure of the solid, especially detailed information on its band structure: the density of states, dispersion relations in bands, effective masses of carriers etc. Therefore the first-principle calculations are dedicated for a particular material.

At high impact energies the scattering rate for electron-hole production in both metals and semiconductors follows Bethe asymptotics. We have checked (not shown) that the rate calculated with Eq.(1) indeed shows the Bethe asymptotic behaviour at high values of the impact energy, $E$, both for metals and semiconductors,

$$\Gamma(E) = (\tilde{A} \ln(E/E_0) + \tilde{B})/\sqrt{E}. \tag{2}$$

The coefficients $\tilde{A}$ and $\tilde{B}$ are material-dependent constants. Their units are: $[\tilde{A}], [\tilde{B}] = eV/\text{Å}$. Coefficient $E_0 = 1$ eV has been introduced for dimensional reasons.

At low energies, $E_G < E < E_P$, the scattering rates obtained with the first-principles calculations are usually well fit with the following formula [15, 18, 34]:

$$\Gamma(E) = \tilde{a}(E - E_{th})^{\tilde{b}}, \tag{3}$$

where $\tilde{a}$ and $\tilde{b}$ are material-dependent coefficients, calculated so as to get $\Gamma$ in [1/fs] units with energy expressed in [eV] units. The effective threshold for pair production, $E_{th}$, may differ from the gap energy, $E_G$. The free-electron-gas model of solids predicts, $\tilde{b} = 2$ [35]. The same value of $\tilde{b} = 2$ may also be obtained with the Fermi-liquid theory [35]. In the literature on semiconductors the quadratic rate is known as the Keldysh scattering rate [36]. It may be obtained analytically from Eq. (1) both for a metal or a direct-band-gap semiconductor, assuming simple parabolic bands and using the constant matrix element approximation (CME) [19]. However, full band calculations show that the value of $\tilde{b}$ approaches 2 only in materials whose structure can be described by the free-electron model (parabolic band structures) [19]. In other solids, $\tilde{b}$ is usually different from 2, and its value reflects the complexity of the band structure.

## 2.3 Impact ionization mean free path, $\lambda_{ii}$, at high energies, $E \gg E_P$

An inelastic collision of a high energy electron, $E \gg E_P$, with a solid may induce either a direct electron-hole production or a plasmon excitation. The energy of the impact electron is then so high, that the interaction is not much affected by the structure of the band in the solid. As the electron's de Broglie wavelength becomes smaller with the increasing energy of the electron impact, electrons of energies high enough will interact only with single atoms. At those high energies, typically a few hundred eVs, the electron-solid interaction may be accurately described by atomic models [3,37,38]. The mean free path can then be described by the Bethe formula [3],

$$\lambda(E) = E/(A \ln(E/E_0) + B), \qquad (4)$$

where $A$ and $B$ are material-dependent constants, related to the coefficients, $\tilde{A}$ and $\tilde{B}$, from Eq. (2). Their units are: $[A], [B] = eV/\text{Å}$. The dimensional coefficient, $E_0 = 1$ eV.

Among atomic and oscillatory models, we found the Gries equation [4], based on the orbital interaction model. This equation gives accurate predictions for electron mean free paths in many elements.

An alternative approach: the free-electron-gas approximation [35] has proved to be very successful in describing the interaction of energetic electrons with solids both at high ($E > 200$ eV) and at intermediate electron energies ($50 < E < 200$ eV). In this approximation, an inelastic collision is modelled as an interaction of a free electron with a gas of non-interacting electrons [31,32]. The mean free path of the impact electron is then inversely proportional to the imaginary part of the electron self-energy. This self-energy depends on the Lindhard dielectric function [39], and may be accurately modelled using experimental data for the optical dielectric function. This is the basis of the optical models [5,6] which are commonly used for calculating electron mean free paths at high and intermediate impact energies. Details of the band structure enter these models only through the shape of the optical dielectric function for a specific solid. Among the optical models, the TPP-2 model and the resulting TPP-2 fit [5,40], provide a comprehensive description of the inelastic electron mean free paths in many elements and compounds over a wide energy range, $50 < E < 2000$ eV. Mean free paths obtained with the TPP-2 model were found to be in a good agreement with the experimental data [41].

We will refer to the results of the TPP-2 model and the Gries model [43] in the forthcoming analysis.

## 2.4 A universal formula

As at low energies, $E_G < E < E_P$, the electron scattering rate, $\Gamma(E)$, shows the scaling behaviour of the form given by Eq. (3), the electron mean free path, $\lambda_{ii}(E)$ may then be estimated as,

$$\lambda_{ii}(E) = \sqrt{E}/(a\,(E - E_{th})^b). \tag{5}$$

Coefficients $a$ and $b$ are material-dependent constants, related to $\tilde{a}$ and $\tilde{b}$ and expressed in the same units as $\tilde{a}$ and $\tilde{b}$.

Since the group velocity and the scattering rate in solids depend both on the wave vector and on the energy band, the correct average of the mean free path is, $\lambda_{ii}(E) \propto \sum_{n,k} \{v_{n,k}(E)/\Gamma_{n,k}(E)\}$. For Eq. (5) we approximate the average by the ratio of averages:

$\lambda_{ii} \propto \left\{ \sum_{n,k} v_{n,k}(E) \right\} / \left\{ \sum_{n,k} \Gamma_{n,k}(E) \right\}$, which is a first-order approximation.

For simplicity we also assume a quadratic dispersion relation of the average velocity and the energy, $v(E) \propto \sqrt{E}$.

For metals, where $E_{th} = 0$, Eq. (5) simplifies to,

$$\lambda_{ii}^{metals}(E) = c\,E^{-d} \tag{6}$$

where $c$ and $d$ are positive coefficients. The value of $d$ calculated with the free-electron-gas model would be: $d = \tilde{b} - 0.5 = 2 - 0.5 = 1.5$. We note that RHSs of both Eqs. (5), (6) strongly decrease as the energy increases. At large energies these RHSs become equal to 0.

At high energies, $E >> E_P$, $\lambda_{ii}$ is well described by the Bethe equation (4). As the energy decreases, the RHS of the Bethe equation decreases faster than $E$. The RHS of Eq. (4) has also an unphysical pole at $E = E_0 \cdot exp(-B/A)$, which changes the sign of $\lambda_{ii}$ at $E < E_0 \cdot exp(-B/A)$ ($E_0 = 1$ eV).

At intermediate energies, when $E$ equals a few times $E_P$, the behaviour of $\lambda_{ii}$ has been investigated experimentally for several elements [2,11,32,42]. The data show a characteristic dip of the $\lambda_{ii}$ curve at energies of a few times $E_P$, and a further increase towards higher energies. At intermediate energies there are no fully reliable theoretical predictions for $\lambda_{ii}$. The exchange and correlation terms in the atomic potential, and the complex structure of the energy bands strongly influence the dynamics of the scattered electrons [5,40]. These effects are not included in the optical models, therefore these models are beyond their validity at those energies. At present, the first-principles calculations which include for these effects underestimate the experimental data at intermediate energies (see e. g. the calculations for Be [2,11]). The authors of Ref. [2], however, expect that this discrepancy may be removed or diminished after improving some parts of their calculations.

Here we propose a simple formula for $\lambda_{ii}$ which takes into account the asymptotics of $\lambda_{ii}$ both at low and at high energies:

$$\lambda_{ii}(E) = \sqrt{E}/(a\,(E - E_{th})^b) + (E - E_0 \cdot exp(-B/A))/(A \ln(E/E_0) + B). \quad (7)$$

For metals, $E_{th} = 0$, and Eq. (7) reduces to:

$$\lambda_{ii}(E) = c\,E^{-d} + (E - E_0 \cdot exp(-B/A))/(A \ln(E/E_0) + B). \quad (8)$$

In both Eqs. , $E_0 = 1$ eV. Eqs. (7), (8) correct the high energy term, $E/(A \ln(E/E_0) + B)$, appearing in (4) by removing the unphysical pole at $E = E_0 \cdot exp(-B/A)$. This does not change the asymptotic behaviour of Eqs. (7), (8), as the $E_0 \cdot exp(-B/A)$ is usually of the order of a few tens of eVs at most, so both equations reduce to (4) at high energies. At low energies the high energy term is small, so the behaviour of $\lambda_{ii}$ is then governed by the low energy term, $\sqrt{E}/(a\,(E - E_{th})^b)$. At intermediate energies both high and low energy terms contribute to $\lambda_{ii}$.

## 3 Mean free path fitted

We fitted Eq. (7) to available data for the impact ionization mean free path for 9 elements and 2 compounds. We have chosen those materials for which either experimental data or phenomenological fits based on the experimental data were available. These materials may be classified into the following groups: (i) alkali metals, (ii) polyvalent metals, (iii) elemental semiconductors and (iv) composite semiconductors.

The coefficients $A$ and $B$ appearing in the high energy term of Eq. (7) were fitted in the following way. As the power-law term in Eq. (7) does not contribute at large energies, we fitted the coefficients $A$, $B$ with the Bethe-like term of these equations, using the high energy data from the NIST database [43] at energies $E > 200$ eV.

Afterwards, we determined the coefficients $a$ and $b$ of the low energy term. We used either the coefficients found in the literature that were obtained with the fits to the first-principles calculations, or we ourselves fitted $a$ and $b$ to the experimental data or to the results of the first-principles calculations. We used the full form of Eq. (7) for the fits, where coefficients, $A$ and $B$, appearing in these equations, were obtained as described in the previous paragraph. Fitting the experimental data, we used the data points recorded at energies up to a few eVs above the plasmon threshold, $E_P$. Figs. 1, 2, 3, 4, 5 show the mean free path for impact ionization obtained with Eq. (7) compared to the predictions of the TPP-2 model ($E > 50$ eV) and the Gries model ($E > 200$ eV) and to the experimental data. The fitting coefficients and their relative standard errors are listed in Tables 1, 2. The coefficients are calculated so as to get $\lambda_{ii}$ in [Å] units with energy expressed in [eV] units.

Mean free paths for 3 alkali metals (Li,Na,K) are plotted in Fig. 1. Alkali metals are monovalent metals usually of the simplest Fermi surfaces [35]. These surfaces are known with a great precision for all alkali metals (Na, K, Rb, Cs), except of Li. They are nearly spherical, and lie entirely inside a single Brillouin zone. Therefore the free-electron-gas models are usually successful in describing the electron transport in alkali metals.

The mean free paths obtained with our fit for alkali metals follow the free-electron results of $b = 1.5$ for Na ($Z = 11$) and K ($Z = 19$). The values of $b$ for these metals are: $b = 1.4$ for Na and $b = 1.9$ for K. For Li ($Z = 3$) we obtain: $b = 3.3$. This may indicate that the Fermi surface of lithium is far from the spherical one.

Mean free paths for Na and Li are well fitted with Eq. (7). However, we observe that the results of the TPP-2 model used for fitting mean free path at high energies underestimate the experimental data for Na at intermediate energies, $50 < E < 100$ eV. This effect is even more pronounced in K ($Z = 19$) and in heavier alkali metals, Rb and Cs (not shown here, see [42]). If this discrepancy can be removed, the accuracy of our fit would improve.

Our next observation is that the mean free path measured experimentally for the alkali metals of higher atomic numbers (K, Rb, Cs) shows a characteristic anomalous dip at low energies [42] which cannot be reproduced with our fit or with the free-electron models. The occurrence of the dip was primarily explained through a possible plasmon excitation followed by the scattering of the conduction electrons into the d-band, which is empty in heavier alkali metals like K, Cs, Rb. This hypotesis was put in question by new experimental results on the electron mean free paths in beryllium ($Z = 4$) [11] which is a divalent metal with no d-bands at all. The electron mean free paths in beryllium also showed the anomalous dip at low energies, which cannot then be explained by a plasmon decay into d-band. The authors of Ref. [11] suggested that the dip might result from some excitations of surface plasmons or of the surface states. Those predictions were based on the free-electron-gas model. A detailed first-principles analysis of the electron mean free paths in Be was afterwards performed in [2]. It included the full band structure of Be. The results obtained were in an agreement with the experimental data recorded at energies, $E < 30$ eV, reproducing the anomalous dip with a good accuracy. This proved again that including the band structure is essential for full understanding of the electron dynamics at low energies. However, at higher energies those first-principles predictions underestimated the experimental data and the results with the free-electron-gas model. The authors of Ref. [2] expected that this discrepancy may be removed or diminished after improving some parts of the calculations.

Results of our fit for beryllium are plotted in Fig. 2. They are compared to the results of the Gries model [4] which in this case was extended down to $E \sim 10$ eV. The low energy data [11] are well fitted by Eq. (7). Our fit agrees better with those data than that with the Gries model. Again, there is a distinct discrepancy between our fit and the experimental data at the intermediate energies, $E = 30 - 70$ eV, when the $\lambda_{ii}$ curve rises from the dip. In this region our fit distinctly underestimates the data.

Mean free paths obtained for other polyvalent metals are plotted in Fig. 3. Both experimental data and first-priciple calculations at low impact energies are available for these metals [7,9,10,32]. Coefficients $a$ and $b$ appearing in Eq. (7), were estimated by fitting to the electron mean free paths obtained with Eq. (5) from the electron lifetimes. Those lifetimes were predicted with first-principles calculations or taken from experiment at very low electron energies: (i) $0.6 < E < 3.2$ eV (calculations) for Al [7], (ii) $1.1 < E < 3.5$ eV (calculations) and $10 < E < 13$ eV (data) for Cu [7,32], and (iii) $0.4 < E < 1.3$ eV (data) for Ag [10]. The values of $b$ obtained from the fits, $\sim 1 - 1.5$, were below the free-electron value of 1.5.

The fits to $\lambda_{ii}$ in polyvalent metals follow the experimental data. The largest discrepancies were again observed at intermediate energies, where data were sparse. More data would be necessary in order to test and improve the accuracy of our fit in this energy regime.

Figs. 4 and 5 show the mean free paths in two elemental semiconductors, Si and C (diamond), and in two composite semiconductors, GaAs and ZnS. At high energies the mean free path was fitted, as previously, to the results of the TPP-2 model. At low energies we used coefficients $a$, $b$ and $E_{th}$ from the existing fits to the scattering rates. Those fits were performed for Si in Refs. [44,45], and for C in Ref. [16,17]. We did the same for the composite semiconductors, using the fitting coefficients found in Refs. [18,20]. The scattering rates in [16–18,20,44,45] were calculated with the first-principles calculations, and then fitted with the power-law equation, Eq. (3).

The accuracies of the first-principles calculations of the scattering rates obtained in Refs. [16–18,20,44,45] were indirectly verified by comparing the predictions obtained with the electron transport model employing these rates to some experimental data. For Si the quantum yield predicted by this model was in a good agreement with the experimental predictions [44]. In Ref. [45] the analysis of the data obtained with the soft-X-ray photoemission spectroscopy, the data on the quantum yield and the ionization coefficients in Si was coupled to the Monte-Carlo model of electron transport and resulted in an empirical expression for the impact-ionization rate in Si as a function of electron energy, $E$, consistent with the data.

Similarly, the scattering rate obtained in Ref. [16, 17] was further applied therein for Monte-Carlo simulations of electron transport in diamond [16, 17]. Those simulations tested the dependence of the drift velocity, obtained with that scattering rate, on the external electric field. The results obtained were in a good agreement with the experimental data.

The scattering rates for GaAs and ZnS from Refs. [18, 20] were also successfully tested in the full-band Monte-Carlo simulations of electron transport under high electric fields.

As we did not find in the literature any direct experimental measurements of the electron mean free paths or lifetimes in Si, C, GaAs or ZnS, we cannot at present compare our predictions to data. The availability of new data recorded at intermediate energies would be of special importance, as this is the region where discrepancies usually appear for metals. This especially applies to diamond, where the predictions of different models are inconsistent even at high energies. Fig. 6 shows the predictions of the TPP-2 model, the TPP-2 fit (Eq. (2) in Ref. [40]), the Gries model and the experimental results on the electron mean free paths in diamond obtained at energies, $E > 200$ eV, from the NIST database [43]. These data were obtained from the experimental data on glassy carbon by density scaling: $\lambda_D = \lambda_C \cdot \rho_D/\rho_C$, where $\rho_D$ was the density of diamond, $3.51\, g/cm^3$, and $\rho_C$ is the density of glassy carbon ($\rho_C = 1.5\, g/cm^3$).

The figure shows large differences among the $\lambda$s predicted, which at $E = 200$ eV extend to $\sim 4$ A and at $E = 2000$ eV extend to $\sim 17$ A. Direct measurements of the mean free paths in diamond would be essential in order to reduce this uncertainity. Considering the low energy part of our fit, we note that it lies much below the predictions of the TPP-2 model at energies $E = 10 - 100$ eV. However, in Ref. [30], we successfully used the TPP-2 model extended down to $E = 9$ eV, and Watanabe's results [16, 17] at energies $5.47 < E < 9$ eV for modelling the energy loss by a single impact electron. We then obtained reasonable values for the total number of cascading electrons liberated by the electron, and the average energy needed to create an electron-hole pair, $E_{eh} = 12 - 12.5$ eV. Those predictions slightly underestimated the experimental values of the average pair-creation energy which have been found between 12.8 and 13.6 eV [46–48] with 13 eV being the most recent result [48].

In order to check how our fit works, we repeated the simulation of electron cascades as in Ref. [30], using Eq. (7) for fitting the electron mean free paths both at high and at low electron energies. We obtained the mean value of pair creation energy of 10.5 eV. This value agrees with the theoretical predictions on cascades initiated by electrons which suggests pair creation values between 10.3 and 11.6 eV [49,50]. However, the value obtained lies somewhat below experimental results [46–48].

The observations made above indicate that the values of mean free paths in diamond obtained with different models are inconsistent. They should be critically reviewed both at high and at the intermediate energies. A new analysis of the electron transport in diamond, employing the predictions revised and maybe our fit, should then be performed and eventually compared to the existing experimental data.

## 4 Discussion and summary

We have proposed a simple equation for fitting the electron mean free paths in solids. This equation includes both low and high energy asymptotics. Using this equation, we fitted electron mean free paths for 9 elements and 2 compounds, including alkali metals, polyvalent metals and elemental and composite semiconductors. The mean free paths calculated with the fits generally follow the experimental data or the first-principles calculations with the exception of the intermediate energy region. In that region our fits do not reproduce the anomalous dips observed in the data, an effect which reflects the complexity of the band structure. Also, in some solids our fit underestimates the rise of the $\lambda$ towards higher energies which occurred after the dip. This discrepancy cannot be corrected by a simple replacement of the Bethe-like term in Eq. (7) with the TPP-2 fit [5]. However, the discrepancies observed were not large - of the order of a few Angstroms at most, and we expect that they will not much affect the average ionization rates.

The fitting equation (7) contained four free parameters for metals, and 5 free parameters for semiconductors. Two (three) of these parameters, $a$, $b$ (and $E_{th}$), appeared in the low-energy term of these equations, and two other parameters, $A$ and $B$, were included in the Bethe-like term of the equations. As in the TPP-2 fit [40], the coefficients $A$ and $B$ may be parametrized as approximate functions of the solid density, $\rho$, of the gap energy, $E_G$, and of the plasmon energy, $E_P$. We do not expect that such parametrization would be possible for coefficients, $a$, $b$ and $E_{th}$, which strongly depend on the details of the band structure of the solid. We expect that any reasonable estimation of these coefficients would require dedicated band calculations.

The fitting equations, Eqs. (7), (8), are similar to the empirical fit by Seah and Dench: $\lambda = \tilde{c}/E^2 + \tilde{d} \cdot E^m$ [51]. However, in our equations the proposed shapes of the low and the high energy terms represent the known asymptotics of electron mean free path, whereas the fit by Seah and Dench [51] is purely empirical. Also, its low energy part cannot be interpreted with the free-electron-gas approximation, since the exponent at $E$ differs from the free electron value.

Our mean free path formula may be used for constructing a general Monte-Carlo model of electron transport, necessary for describing the interaction of the energetic photons with matter. The differential cross sections needed for estimating the energy loss by impact electrons may be approximated by the Bethe differential cross section at high electron energies and the loss probability based on the random-k approximation [13] at low energies. Monte Carlo codes employing Eq. (7) and the approximate differential cross sections, would then be efficient computationally. Such models may be applied to simulate ionization phenomena induced by electrons released by the FEL photons in many different systems, ranging from the explosion of atomic clusters to the formation of warm dense matter and plasmas. The model can also be used to estimate ionization rates and the spatio-temporal characteristics of secondary electron cascades in biological substances. In general, we expect that such models may be used in all applications in which it is important to follow the ionizations by an impact electron down to its very low impact energies. However, in the FEL applications, as the number of electrons released within the sample by photons from the FEL will be large, the number of ions in the sample will increase rapidly. Therefore one will need to correct the mean free path formula including the effect of prior ionization.

In summary, we propose an effective method of fitting electron mean free paths for impact ionization in solids over a wide energy range. Our fit is based on: (i) the results of first-principles calculations or experimental data available at low energies, and (ii) on the results of the well-established optical model at high energies. We have presented a detailed review of the available data and first-principles calculations on $\lambda_{ii}$ in 9 elements and 2 compounds. The derived fitting curves were then compared to these data, in order to test the accuracy of the fit. The largest discrepancies were observed in the region of intermediate energies, where both low and high energy terms contribute to the mean free path. As the effects of complex band structure of the solids are strongly manifested also in this energy regime, the free-electron-gas models were found to be no longer valid at intermediate energies. The available first-principles calculations, however, underestimate the data in this region. We expect that our fit which joins both the high and low energy approach, will be of use until first-principles calculations are sucessfully extended to the intermediate energies, and parametrized there. They can then be directly linked to the optical models at high energies.

The formulae and fitting coefficients developed in this paper show that the region of low energies is no longer inaccesible for the analysis of electron mean free paths. Therefore a comprehensive description of an electron mean free path should also include a description of $\lambda$ at low energies. We believe that our fit gives a useful approximation for $\lambda_{ii}$ both at high and at low energies, and that it will inspire searching for more accurate and universal methods of fitting the mean free paths at the wide energy range.

# Acknowledgments

We are grateful to E. Weckert and I. Vartaniants for reading the manuscript and useful comments. This research was supported in part by the Polish Committee for Scientific Research with grant No. 2 P03B 02724. Beata Ziaja is a fellow of the Alexander von Humboldt Foundation.

# References

[1] C. J. Powell and A. Jablonski. *J. Vac. Sci. Technol. A*, 17 (4):1122, 1999.

[2] E. V. Chulkov V. M. Silkin and P. M. Echenique. *Phys. Rev. B*, 68:205106, 2003.

[3] H. Bethe. *Ann. Phys. (Leipzig)*, 5:325, 1930.

[4] W. H. Gries. *Surf. Interface Anal.*, 24:38, 1996.

[5] S. Tanuma, C.J. Powell, and D.R. Penn. *Surf. Interf. Anal.*, 11:577, 1988.

[6] J.C. Ashley. *J. Elec. Spec. Rel. Phenom.*, 50:323–334, 1990.

[7] I. Campillo et al. *Phys. Rev. B*, 61:13486, 2000.

[8] W. D. Schoene, R. Keyling, M. Bandic, and W. Ekardt. *Phys. Rev. B*, 60:8616, 1999.

[9] V. P. Zhukov and E. V. Chulkov. *J. Phys. Condens. Matter*, 14:1937, 2002.

[10] R. Keyling, W. D. Schoene, and W. Ekardt. *Phys. Rev. B*, 61:1670, 2000.

[11] L. I. Johansson and B. E. Sernelius. *Phys. Rev. B*, 50:16817, 1994.

[12] A. Fleschar and W. Hanke. *Phys. Rev. B*, 56:10228, 1997.

[13] E. O. Kane. *Phys. Rev.*, 159:624, 1967.

[14] IBM Research. Damocles home page. *http://www.research.ibm.com/DAMOCLES*.

[15] T. Kunikiyo et al. *J. Appl. Phys.*, 79:7718, 1996.

[16] T. Watanabe et al. *Jpn. J. Appl. Phys. 2*, 40(7B):L715, 2001.

[17] T. Watanabe et al. *J. Appl. Phys.*, 95:4866, 2004.

[18] M. Reigrotzki et al. *J. Appl. Phys.*, 86:4458, 1999.

[19] N. Sano and A. Yoshii. *J. Appl. Phys.*, 77:2020, 1995.

[20] H. K. Jung, H. Nakano, and K. Taniguchi. *Physica B*, 272:244, 1999.

[21] R. Henderson. *Proc. R. Soc. Lond. Biol. Sci.*, 241:6–8, 1990.

[22] R. Henderson. *Q. Rev. Biophys.*, 28:171–193, 1995.

[23] D. Sayre and H. N. Chapman. *Acta Cryst. A*, 51:237, 1995.

[24] R. Neutze, R. Wouts, D. van der Spoel, E. Weckert, and J. Hajdu. *Nature*, 406:752–757, 2000.

[25] R. A. London, R. M. Bionta, R. O. Tatchyn, and S. Roesler. Optics for Fourth-Generation X-Ray Sources. *Proc. SPIE*, 4500:51, 2001.

[26] R. W. Lee et al. *Laser and Particle Beams*, 20:527, 2002.

[27] K. Sokolowski-Tinten and D. von der Linde. *Phys. Rev. B.*, 61:2643, 2000.

[28] H. Wabnitz et al. *Nature*, 420:482, 2002.

[29] B. Ziaja, A. Szöke, D. van der Spoel, and J. Hajdu. *Phys. Rev. B*, 66:024116, 2002.

[30] B. Ziaja, R. London, and J. Hajdu. *J. Appl. Phys.*, 97:064905, 2005.

[31] J. L. Quinn. *Phys. Rev.*, 126:1453, 1962.

[32] D. R. Penn. *Phys. Rev. B*, 35:482, 1987.

[33] S. Samarin et al. *Surf. Sci.*, 548:187, 2004.

[34] H. K. Jung, K. Taniguchi, and C. Hamaguchi. *J. Appl. Phys.*, 79:2473, 1996.

[35] N.W. Ashcroft and N.D. Mermin. Solid state physics. *Harcourt, Inc.*, 1976.

[36] L. V. Keldysh. *Sov. Phys. JETP*, 21:1135, 1965.

[37] Y.-K. Kim and M.E. Rudd. *Phys. Rev. A*, 50:3954, 1994.

[38] Y.-K. Kim, J.P. Santos, and F. Parente. *Phys. Rev. A*, 62:052710, 2000.

[39] J. Lindhard. *K. Dan. Vidensk. Selsk. Mat. Fys. Medd.*, 28:1–57, 1954.

[40] S. Tanuma, C.J. Powell, and D.R. Penn. *Surf. Interf. Anal.*, 17:911, 1991.

[41] W. S. M. Werner et al. *J. Elec. Spec. Rel. Phenom.*, 113:127, 2001.

[42] N. V. Smith, G. K. Wertheim, A. B. Andrews, and C. T. Chen. *Surf. Sci. Lett.*, 282:L359, 1993.

[43] A. Jablonski and C. Powell. NIST Electron Inelastic Mean Free Paths. NIST Standard Reference Database 71, 2000.

[44] Y. Kamakura et al. J. Appl. Phys. 88:5802, 2000.

[45] E. Cartier, M. V. Fischetti, E. A. Eklund, and F. R. McFeely. *Appl. Phys. Lett.*, 62:3339, 1993.

[46] S. F. Kozlov, R. Stuck, M. Hage-Ali, and P. Siffert. *IEEE Trans. Nucl. Sci*, NS-22:160, 1975.

[47] C. Canali et al. *Nucl. Inst. Meth.*, 160:73, 1979.

[48] F. Nava et al. *IEEE Trans. Nuc. Sci.*, NS-26:308, 1979.

[49] R. C. Alig, S. Bloom, and C. W. Struck. *Phys. Rev. B*, 22:5565, 1980.

[50] R. C. Alig. *Phys. Rev. B*, 27:968, 1983.

[51] M. P. Seah and W. A. Dench. *Surf. Interface Anal.*, 1:2, 1979.

# Figure captions

**Fig. 1.** Electron mean free path in 3 alkali metals: (a) lithium, (b) sodium, and (c) potassium, fitted with Eq. (8) (solid line) and compared to the predictions of the TPP-2 (dashed line) and the Gries (dash-dotted line) models and to the experimental data (circles). Thin dashed lines show separate contributions of the low-energy and the high-energy asymptotics of Eq. (8).

**Fig. 2.** Electron mean free path in beryllium fitted with Eq. (8) (solid line) and compared to the predictions of the TPP-2 (dashed line) and the Gries (dash-dotted line) models and to the experimental data (circles)[11]. Thin dashed lines show separate contributions of the low-energy and the high-energy asymptotics of Eq. (8).

**Fig. 3.** Electron mean free path in polyvalent metals: (a) aluminium, (b) copper, and (c) silver, fitted with Eq. (8) (solid line) and compared to the predictions of the TPP-2 (dashed line) and the Gries (dash-dotted line) models and to the experimental data (filled circles, filled triangles and open triangles). Thin dashed lines show separate contributions of the low-energy and the high-energy asymptotics of Eq. (8).

**Fig. 4.** Electron mean free path in two elemental semiconductors: (a) silicon, and (b) diamond, fitted with Eq. (7) (solid line) and compared to the predictions of the TPP-2 (dashed line) and the Gries (dash-dotted line) models and to the results of first-principles calculations by Kamakura[44] and Watanabe[17] (dotted line with crosses). Thin dashed line shows separate contribution of the high-energy asymptotics of Eq. (7).

**Fig. 5.** Electron mean free path in two composite semiconductors: (a) GaAs, and (b) ZnS, fitted with Eq. (7) (solid line) and compared to the predictions of the TPP-2 (dashed line) and the Gries (dash-dotted line) models and to the results of first-principles calculations by Jung[34] and Reigrotzki[18] (dotted line with crosses). Thin dashed line shows separate contribution of the high-energy asymptotics of Eq. (7).

**Fig. 6.** Electron mean free path of electron in diamond at high energies, $E > 200$ eV, obtained with the TPP-2 model (solid line), the Gries model (dash-dotted line) and the TPP-2 fit (dashed line). The experimental data shown in the figure (dotted line) were scaled from the data on glassy carbon ($\rho_C = 1.5\ g/cm^3$) obtained by Lesiak[43].

# Table captions

**Tab. 1.** Fitting coefficients and their relative standard errors (%) for metals: Li, Na, K, Be, Al, Cu, Ag. The units are: $[A], [B] = eV/\text{Å}$. Coefficients $c$ and $d$ in Eq. (8) are calculated so as to get $\lambda_{ii}$ in Angstroms with energy expressed in [eV] units.

**Tab. 2.** Fitting coefficients and their relative standard errors (%) for semiconductors: Si, C (diamond), GaAs, ZnS. Coefficients $a$, $b$ and $E_{th}$ were taken from the literature [15–18]. The units are: $[A], [B] = eV/\text{Å}$. Coefficients $a$ and $b$ in Eq. (7) are calculated so as to get $\lambda_{ii}$ in Angstroms with energy expressed in [eV] units.

|    | A            | B             | $\log c$   | d         |
|----|--------------|---------------|------------|-----------|
| Li | 4.49(0.07%)  | -5.9(0.4%)    | 8.9(7%)    | 3.3(7%)   |
| Na | 4.85(0.08%)  | -8.6(0.3%)    | 4.4(10%)   | 1.4(13%)  |
| K  | 3.35(0.07%)  | -4.8(0.3%)    | 4(34%)     | 1.9(40%)  |
| Be | 10.88(0.07%) | -25.1(0.2%)   | 14(12%)    | 4.7(12%)  |
| Al | 11.51(0.04%) | -28.6(0.1%)   | 5.18(1%)   | 1.21(7%)  |
| Cu | 17.18(0.07%) | -57.6(0.1%)   | 5.83(0.8%) | 1.48(2%)  |
| Ag | 19.74(0.1%)  | -6.6(0.3%)    | 5.53(0.9%) | 1.00(9%)  |

Table 1:

|      | A            | B              | a                  | b    | $E_{th}$ [eV] |
|------|--------------|----------------|--------------------|------|---------------|
| Si   | 9.26(0.03%)  | -23.20(0.09%)  | $10^{-5}$          | 4.6  | 1.1           |
| C    | 14.0(2.5%)   | -38(8%)        | $4.78\cdot10^{-6}$ | 4.5  | 5.47          |
| GaAs | 8.11(0.5%)   | -27.3(1%)      | $8.17\cdot10^{-5}$ | 4.44 | 1.73          |
| ZnS  | 9.74(0.04%)  | -28.91(0.09%)  | $9.89\cdot10^{-6}$ | 5.07 | 3.8           |

Table 2:

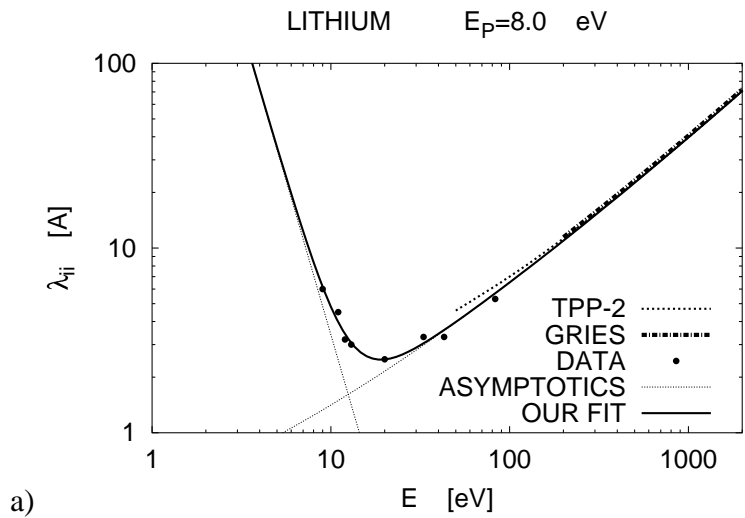

a)

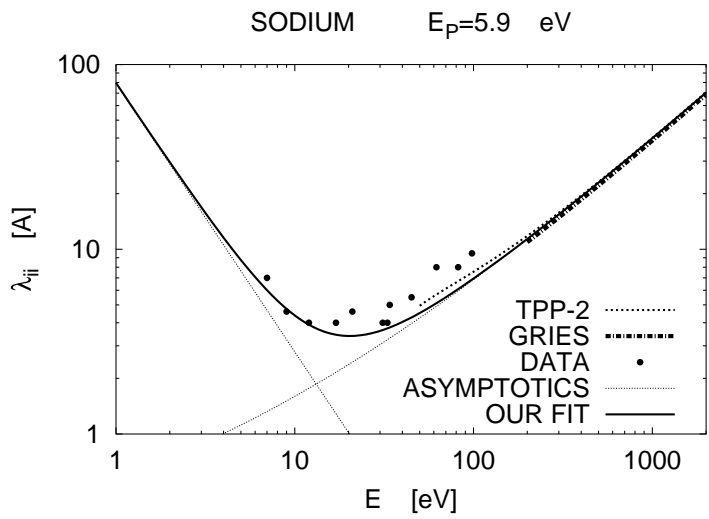

b)

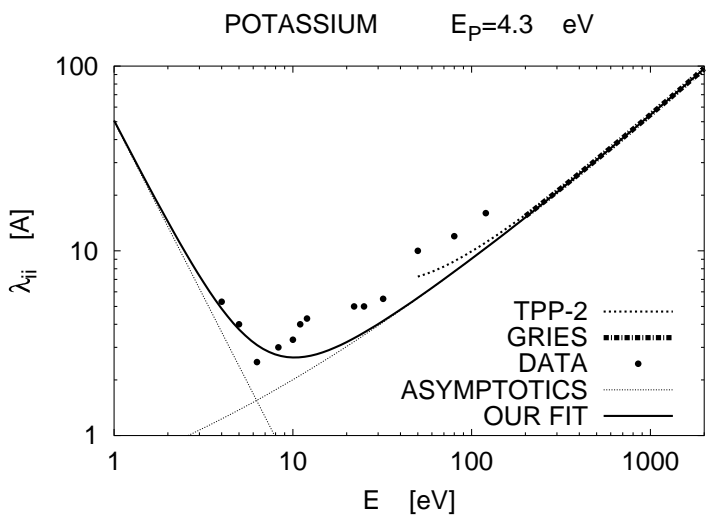

c)

Figure 1:

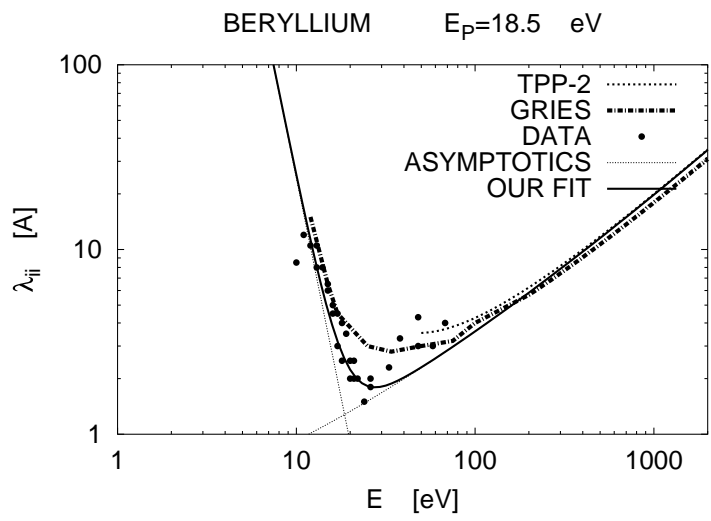

Figure 2:

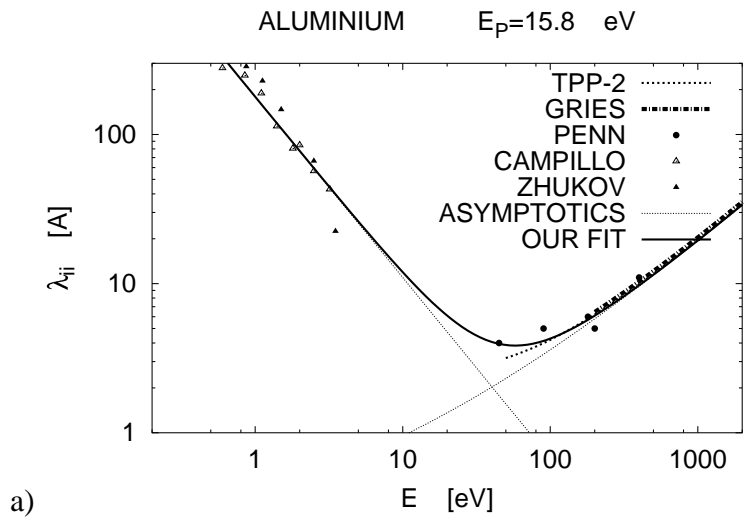

a)

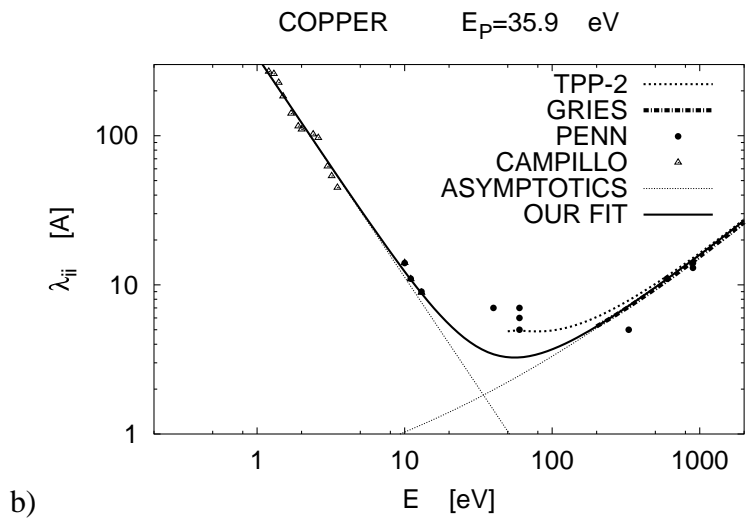

b)

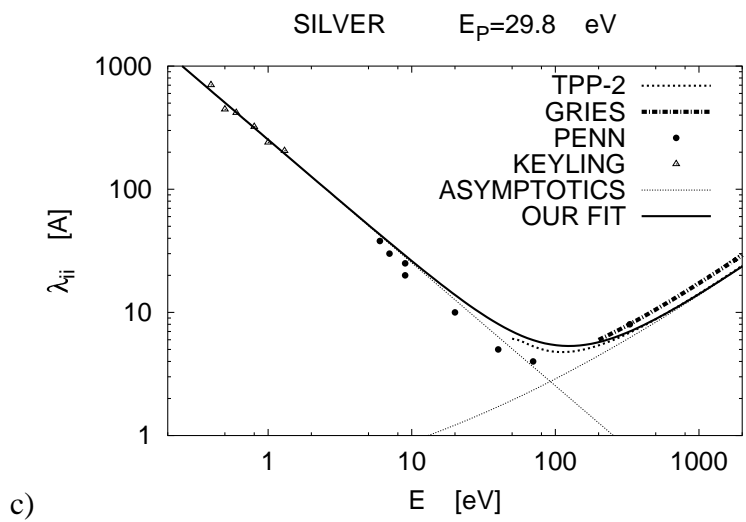

c)

Figure 3:

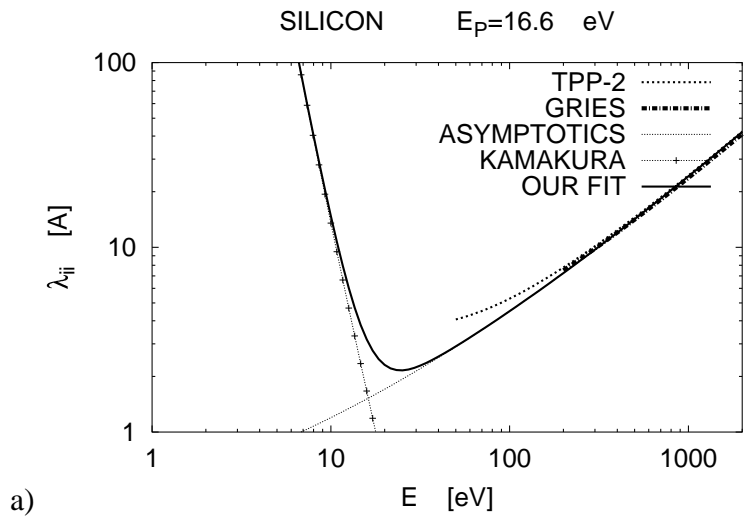

a)

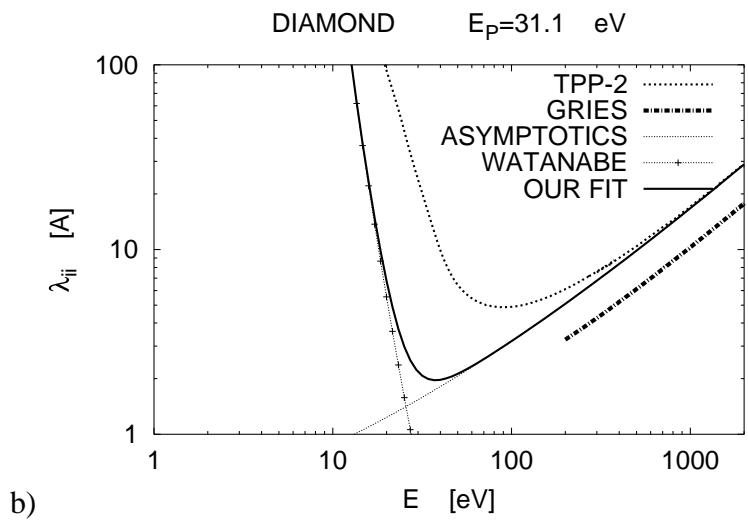

b)

Figure 4:

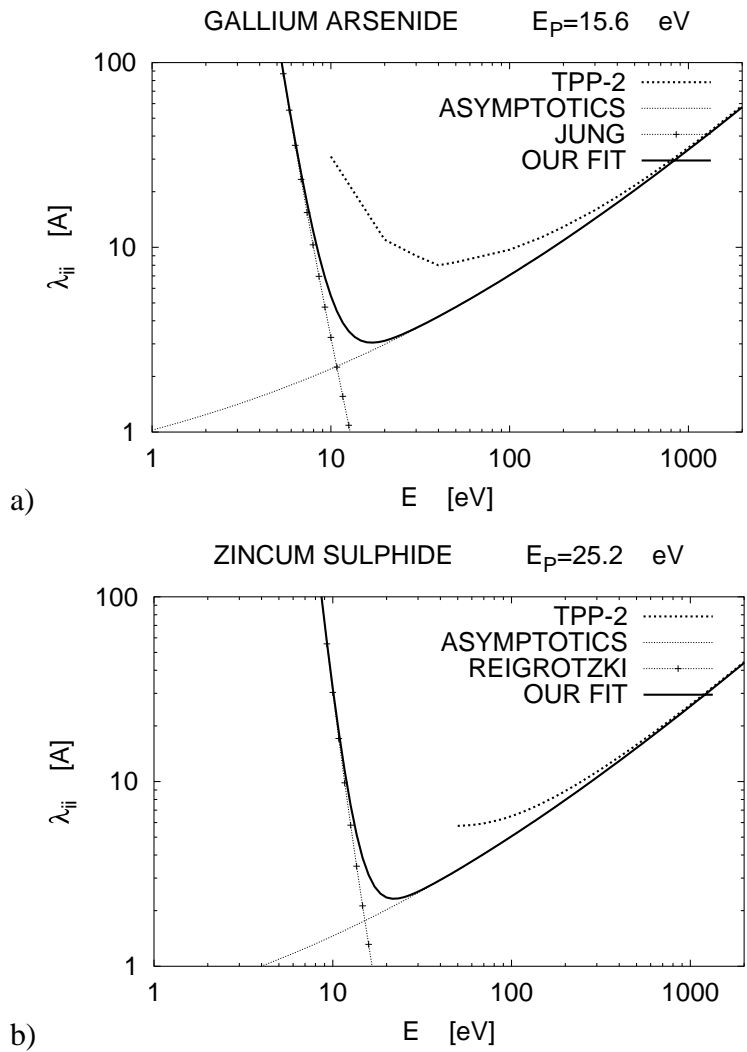

a)

b)

Figure 5:

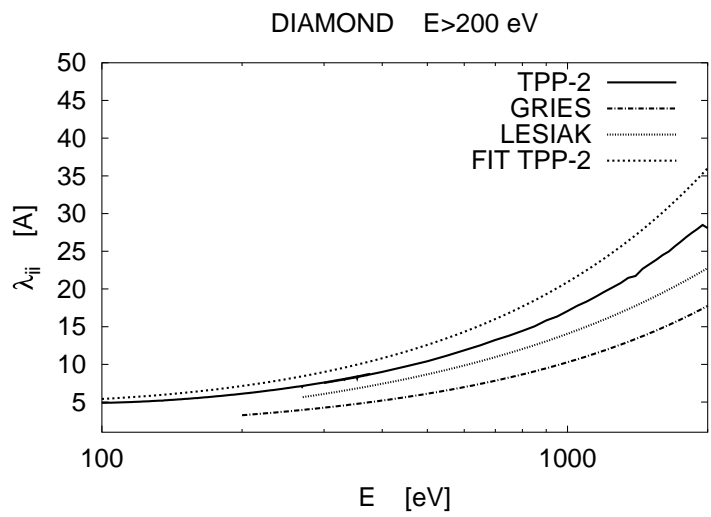

Figure 6: